\RequirePackage{fix-cm}
\documentclass[smallextended]{svjour3}       
\smartqed  
\usepackage{graphicx}
\begin{document}

\title{Precise measurement of positronium hyperfine splitting using the Zeeman effect
}


\author{Akira Ishida         \and
        Yuichi Sasaki        \and 
        Ginga Akimoto        \and 
        Taikan Suehara        \and 
        Toshio Namba        \and 
        Shoji Asai        \and 
        Tomio Kobayashi        \and 
        Haruo Saito        \and 
        Mitsuhiro Yoshida        \and 
        Kenichi Tanaka        \and 
        Akira Yamamoto
}


\institute{A. Ishida \at
              Department of Physics, Graduate School of Science, and International Center for Elementary Particle Physics (ICEPP), The University of Tokyo, 7-3-1 Hongo, Bunkyo-ku, Tokyo 113-0033, Japan \\
              Tel.: +81-3-3815-8384\\
              Fax: +81-3-3814-8806\\
              \email{ishida@icepp.s.u-tokyo.ac.jp}           
           \and
           Y. Sasaki \and G. Akimoto \and T. Suehara \and T. Namba \and S. Asai \and T. Kobayashi \at
              Department of Physics, Graduate School of Science, and International Center for Elementary Particle Physics (ICEPP), The University of Tokyo, 7-3-1 Hongo, Bunkyo-ku, Tokyo 113-0033, Japan
           \and
           H. Saito \at
              Department of General Systems Studies, Graduate School of Arts and Sciences, The University of Tokyo, 3-8-1 Komaba, Meguro-ku, Tokyo 153-8902, Japan
           \and
           M. Yoshida \and K. Tanaka \and A. Yamamoto \at
              High Energy Accelerator Research Organization (KEK), 1-1 Oho, Tsukuba, Ibaraki 305-0801, Japan
}

\date{Received: date / Accepted: date}

\maketitle

\begin{abstract}
Positronium is an ideal system for the research of the quantum electrodynamics (QED) in bound state. The hyperfine splitting (HFS) of positronium, $\Delta _{\mathrm{HFS}}$, gives a good test of the bound state calculations and probes new physics beyond the Standard Model. A new method of QED calculations has revealed the discrepancy by 15\,ppm (3.9$\sigma$) of $\Delta _{\mathrm{HFS}}$ between the QED prediction and the experimental average. There would be possibility of new physics or common systematic uncertainties in the previous all experiments. We describe a new experiment to reduce possible systematic uncertainties and will provide an independent check of the discrepancy. We are now taking data and the current result of $\Delta _{\mathrm{HFS}} = 203.395\,1 \pm 0.002\,4 (\mathrm{stat.}, 12\,\mathrm{ppm}) \pm 0.001\,9 (\mathrm{sys.}, 9.5\,\mathrm{ppm})\,\mathrm{GHz} $ has been obtained so far. A measurement with a precision of $O$(ppm) is expected within a year.
\end{abstract}

\section{Introduction}
\label{intro}
Positronium (Ps), a bound state of an electron and a positron, is a purely leptonic system 
which allows for very sensitive tests of Quantum ElectroDynamics (QED). 
The precise measurement of the hyperfine splitting between orthopositronium (o-Ps, 1$^3S_1$) and parapositronium (p-Ps, 1$^1S_0$) 
(Ps-HFS, $\Delta _{\mathrm{HFS}}$) provides a good test of bound state QED. 
Ps-HFS is expected to be relatively large (for example compared to hydrogen HFS) due to a relatively large 
spin-spin interaction, and also due to the contribution from vacuum oscillation 
(o-Ps $ \rightarrow \gamma ^{\ast} \rightarrow$ o-Ps). 
The contribution from vacuum oscillation is sensitive to new physics beyond the Standard Model. 

\begin{figure}[htbp]
\includegraphics[width=0.5\textwidth]{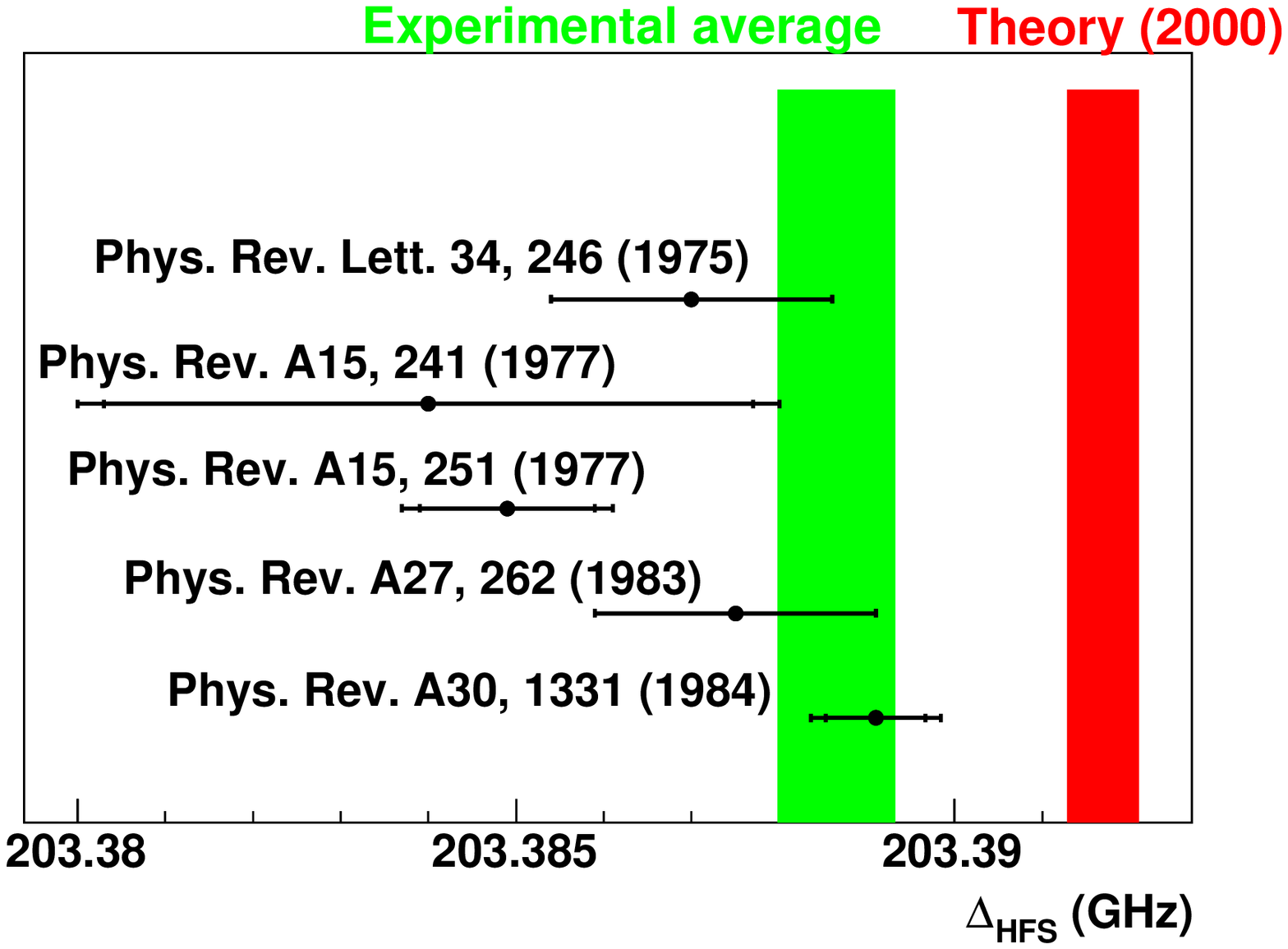}
\caption{Historical plot of $\Delta _{\mathrm{HFS}}$.}
\label{fig:history}
\end{figure}

Fig.~\ref{fig:history} shows the measured and theoretical values of $\Delta _{\mathrm{HFS}}$.
The combined value from the results of the independent and the most precise 2 experiments is 
$\Delta _{\mathrm{HFS}} ^{\mathrm{exp}} = 203.388\,65(67)\,\mathrm{GHz} \,(3.3\,\mathrm{ppm})$
~\cite{HUGHES-V,MILLS-I,MILLS-II}. 
Recent developments in NonRelativistic QED (NRQED) have added 
$O(\alpha ^3 \ln \alpha)$ corrections to the theoretical prediction which now stands at 
$\Delta_{\mathrm{HFS}} ^{\mathrm{th}} = 203.391\,69(41)\,\mathrm{GHz}\,(2.0\,\mathrm{ppm})$~\cite{HFS-ORDER3-KNIEHL,HFS-ORDER3-MELNIKOV,HFS-ORDER3-HILL}. 
The discrepancy of 3.04(79)\,MHz (15\,ppm, 3.9$\sigma$) between $\Delta _{\mathrm{HFS}} ^{\mathrm{exp}}$ 
and $\Delta_{\mathrm{HFS}} ^{\mathrm{th}}$ might 
either be due to the common systematic uncertainties in the previous experiments or to 
new physics beyond the Standard Model. 

There are two possible common systematic uncertainties in the previous experiments. 
One is the unthermalized o-Ps contribution which results in an underestimation of 
the material effect~\cite{ISHIDA_ARXIV}. 
This effect has already been shown to be significant~\cite{KATAOKA,KATAOKA-D,JINNAI,JINNAI-D,ASAI,ASAI-D} in the 
o-Ps lifetime puzzle. 
Another is the uncertainty in the magnetic field uniformity which was cited as 
the most significant systematic error in the previous experiments. 

\begin{figure}
\includegraphics[width=0.35\textwidth]{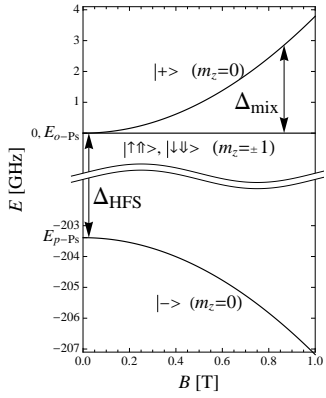}
\caption{Zeeman energy levels of Ps in its ground state. The arrows $\uparrow, \downarrow$ means 
the spin up and down of electron, and the arrows $\Uparrow, \Downarrow$ means the spin up and down of positron.}
\label{fig:pslevel}
\end{figure}

The energy levels of the ground state of Ps are shown as a function of static magnetic field 
in Fig.~\ref{fig:pslevel}. 
Due to technical difficulties in directly stimulating $\Delta _{\mathrm{HFS}}$, 
the previous experiments were indirect measurement by stimulating the transition of 
Zeeman splitting ($\Delta _{\mathrm{mix}}$) under static magnetic fields.
The relationship between $\Delta _{\mathrm{HFS}}$ and $\Delta _{\mathrm{mix}}$ is 
approximately given by the Breit-Rabi equation 
\begin{equation}
\Delta _{\mathrm{mix}} \simeq \frac{1}{2} \Delta _{\mathrm{HFS}} \left( \sqrt{1+4x^2} - 1 \right) \, .
\label{eq:Zeeman}
\end{equation}
$x$ is given as $g^{\prime}\mu _B B  / \left( h\Delta _{\mathrm{HFS}} \right)$, 
where $g^{\prime} = g\left( 1-\frac{5}{24} \alpha ^2 \right)$ 
is the $g$ factor for a positron (electron) in Ps~\cite{G-FACTOR-1,G-FACTOR-2,G-FACTOR-3,G-FACTOR-4}, 
$\mu _B$ is the Bohr magneton, $B$ is the static magnetic field, and $h$ is the Planck constant. 
Briefly, $\Delta _{\mathrm{mix}}$ is measured by applying microwaves under the static magnetic field 
and then $\Delta _{\mathrm{HFS}}$ is obtained from Eq.~(\ref{eq:Zeeman}).
The experimental signal is the change of 2$\gamma$ and 3$\gamma$ annihilation rate caused by Zeeman transition.
The experimental resonance line shape is obtained as explained in the ref.~\cite{ISHIDA_ARXIV}.

\section{Experimental Setup}
\label{sec:experimentalsetup}
\begin{figure}
\includegraphics[width=0.4\textwidth]{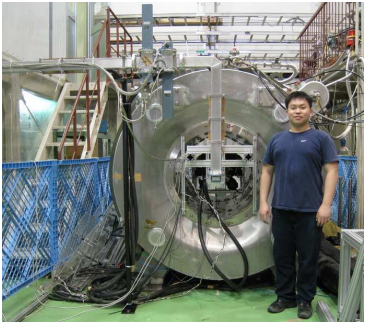}
\caption{Photograph of the setup of our new experiment. The large bore superconducting magnet and the microwave waveguide are shown. Microwaves are guided through the waveguide into the cavity. The RF cavity, $\beta$-tagging system, and the $\gamma$-ray detectors are located at the center of the magnet.}
\label{fig:photo}
\end{figure}

\begin{figure}
\includegraphics[width=1.2\textwidth]{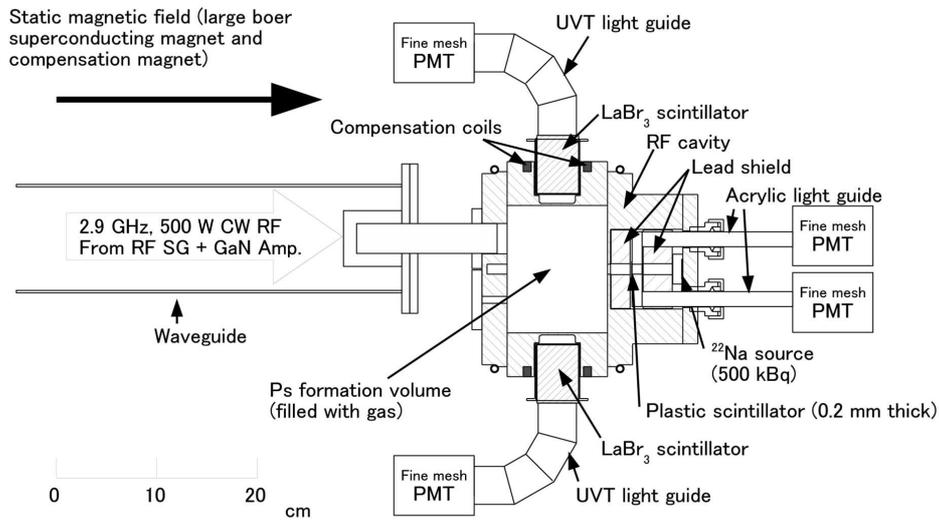}
\caption{Schematic diagram of the setup (top view in magnet).}
\label{fig:schematic}
\end{figure}

The photograph of our new experimental setup is shown 
in Fig.~\ref{fig:photo} and the 
schematic diagram of the setup is shown in Fig.~\ref{fig:schematic}.
The characteristic features of our experiment are the large bore superconducting magnet, 
$\beta$-tagging system, and high performance $\gamma$-ray detectors. 
Details are discussed in the following sections. 

\subsection{Large bore superconducting magnet}
\label{sec:largeboresuperconductingmagnet}
A large bore superconducting magnet is 
used to produce the static magnetic field $B \sim 0.866\,\mathrm{T}$, which induces the Zeeman splitting. 
The bore diameter of the magnet is 800\,mm, and its length is 2\,m. 
The magnet is operated in persistent current mode, making the stability of the 
magnetic field better than $\pm$1\,ppm.
With the compensation coils, the uniformity of the magnetic field is 0.9\,ppm (RMS) 
in the large volume of cylinder 40\,mm in diameter and 100\,mm long, where Ps are formed.

\subsection{$\beta$-tagging system}
\label{sec:betataggingsystem}
The positron source is 14\,$\mu$Ci (500\,kBq) of $ \mathrm{^{22}Na}$.
A plastic scintillator 10\,mm in diameter and 0.2\,mm thick is used to tag 
positrons emitted from the $\mathrm{^{22}Na}$.
The scintillation light is detected by fine mesh PMTs 
and provides a start signal which corresponds to the time of Ps formation. 
The positron stop in the microwave cavity, filled with i-C$_{4}$H$_{10}$ gas, forming Ps.

Ps decays into photons that are detected with LaBr$_3$\,(Ce) scintillators. 
The timing difference of positron emission and $\gamma$-detection is 
used to improve the accuracy of the measurement of $\Delta _{\mathrm{HFS}}$ as follows: 
\begin{enumerate}
\item Imposing a time cut means that we can select well thermalized Ps, reducing the unthermalized o-Ps contribution. 
\item A time cut also allows us to remove the annihilation events (prompt peak in Fig.~\ref{fig:time_and_energy}(a)).
The S/N is improved significantly by about a factor of 20. 
\end{enumerate}

\subsection{High performance $\gamma$-ray detectors}
\label{sec:highperformancegammaraydetectors}
Six $\gamma$-ray detectors are located around the microwave cavity to detect 
the 511\,keV annihilation $\gamma$-rays. 
LaBr$_3$ scintillators, 1.5 inches in diameter and 2 inches long are used. 
LaBr$_3$ scintillators have 
good energy resolution (4\% FWHM at 511\,keV) and timing resolution (0.2\,ns FWHM at 511\,keV), 
and have a short decay constant (16\,ns). 

\section{Analysis}
The experiment was performed from July 2010 to 11th March 2011. Currently the run was stopped by the 
earthquake occurred on 11th March 2011, but will be restarted soon.
In the overall periods, the trigger rate was about 2.4\,kHz and the data acquisition rate was about 
1.2\,kHz. The data acquisition 
was performed using NIM and CAMAC systems. 

\begin{figure}
\includegraphics[width=1.\textwidth]{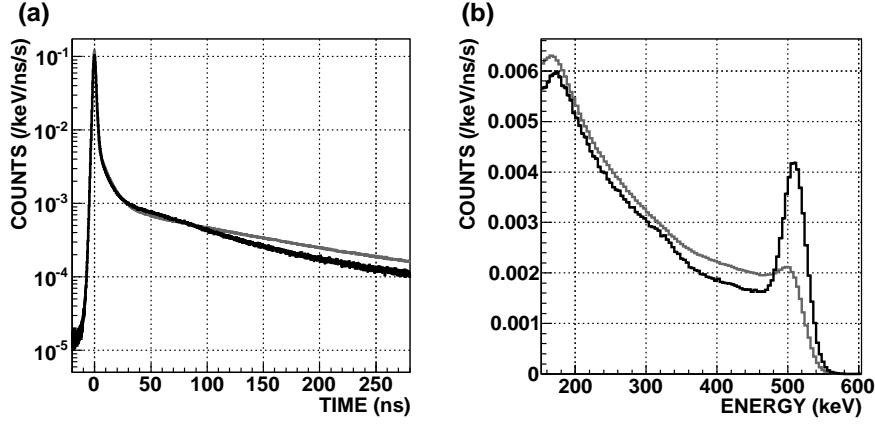}
\caption{Timing and energy spectra. The black lines are RF-ON, and the gray lines are RF-OFF. (a) Decay curves of Ps. The decay rate of Ps increases with RF because of the Zeeman transition. (b) Energy spectra.}
\label{fig:time_and_energy}
\end{figure}

Fig.~\ref{fig:time_and_energy}(a) shows one of the measured timing spectra. The prompt peak contains the 
two processes, annihilation and $|-\rangle$ decay. Decay curve of $|+\rangle$ and 
$m_z = \pm 1$ states contribute dominantly after 30\,ns, 
then the constant accidental spectrum. A timing window of 35--155\,ns 
is applied to select transition events. Fig.~\ref{fig:time_and_energy}(b) shows 
the energy spectra after the timing cut. It is obtained by 
subtracting the accidental contribution using the timing window of 950--1350\,ns.

\begin{figure}
\includegraphics[width=0.5\textwidth]{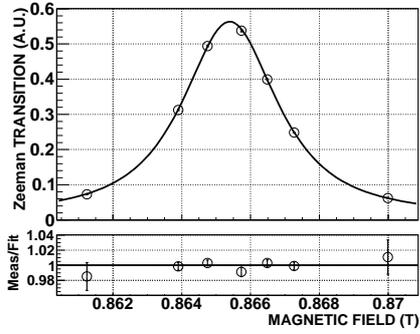}
\caption{Resonance line at 0.895\,15(20)\,amagat gas density. The circles and error bars are the data, and the solid line is the best fit result. The lower figure shows the ratio of measured data over best-fit function. The error bars include errors from statistics of data, statistics of Monte Carlo simulation, uncertainty of RF power, and uncertainty of $Q_L$ value of the RF cavity.}
\label{fig:resonance}
\end{figure}

The difference between RF-ON spectrum and RF-OFF spectrum is the Zeeman transition amount essentially.
The detection efficiency which is necessary to obtain the transition amount 
is estimated by Monte Carlo simulation. 
The 2$\gamma$ and 3$\gamma$ decay rate 
were measured at various magnetic field strengths with a fixed RF frequency as shown 
in Fig.~\ref{fig:resonance}.
Fig.~\ref{fig:resonance} shows one of the resonance lines 
we obtained. The data are fitted by theoretical resonance line. 
To estimate the material effect, the resonance curves have been measured at four gas 
densities. The fitting results are summarized in Table~\ref{tab:result}.

\begin{table}
\caption{Current fitting results of the resonance lines. These uncertainties include errors from statistics of data, statistics of Monte Carlo simulation, uncertainty of RF power, and uncertainty of $Q_L$ value of the RF cavity.}
\label{tab:result}
\begin{tabular}{rrrr}
\hline \noalign{\smallskip}
Gas density (amagat) & $\Delta_{\mathrm{HFS}}$ (GHz) & Relative error (ppm) & $\chi ^{2}/\mathrm{ndf}$ ($p$) \\
\noalign{\smallskip}\hline\noalign{\smallskip}
0.895\,15(20) & 203.350\,6(20) & 9.8 & 4.1/5 (0.54)\\
0.673\,99(19) & 203.359\,3(25) & 12 & 6.7/4 (0.15)\\
0.234\,91(11) & 203.384\,6(24) & 12 & 3.7/3 (0.30)\\
0.168\,434(21) & 203.383\,3(51) & 25 & 2.7/4 (0.61)\\
\noalign{\smallskip}\hline\noalign{\smallskip}
\end{tabular}
\end{table}

\begin{figure}
\includegraphics[width=0.5\textwidth]{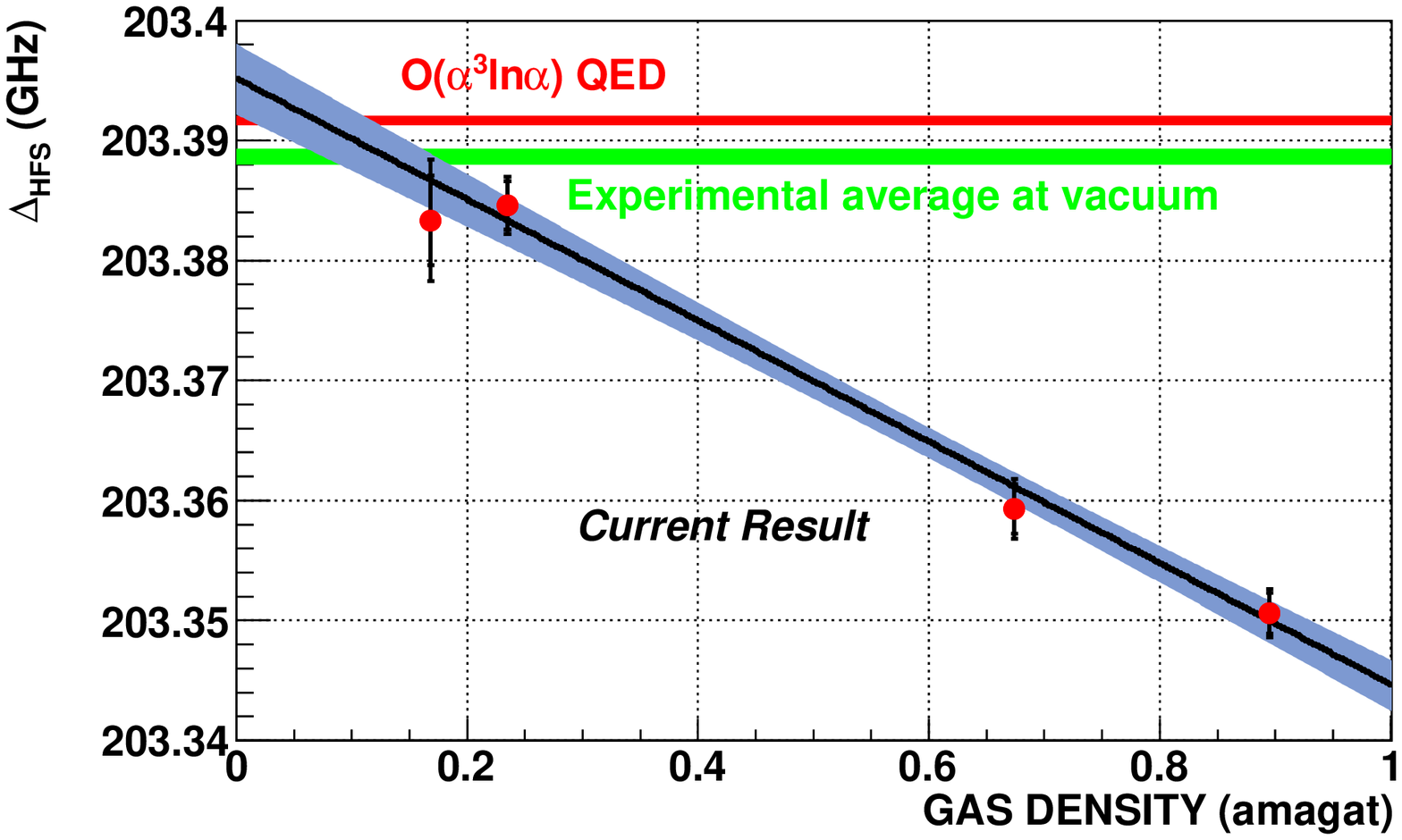}
\caption{Gas density dependence of $\Delta _{\mathrm{HFS}}$. The circles and error bars are the data, the solid line is the best-fit with linear function, the blue band is 1$\sigma$ uncertainty, the green band is experimental average at vacuum, and the red band is $O(\alpha^{3}\ln \alpha )$ QED prediction.}
\label{fig:gas}
\end{figure}

The gas density dependence of $\Delta_{\mathrm{HFS}}$ is shown in Fig.~\ref{fig:gas}.
Currently $\Delta_{\mathrm{HFS}}$ at vacuum is obtained by extrapolating the data linearly.
The fitting results in $\Delta_{\mathrm{HFS}} = 203.395\,1 \pm 0.002\,9 \,\mathrm{GHz}\,(14\,\mathrm{ppm})$
and the gas density dependence is $-248 \pm 21 \,\mathrm{ppm/amagat}$.
The error includes statistical uncertainty, uncertainty from statistics of Monte Carlo simulation, uncertainty of RF power, 
and uncertainty of $Q_L$ value of the RF cavity.
Systematic errors of the current result are summarized in Table~\ref{tab:systematicerrors}.

\begin{figure}
\includegraphics[width=0.5\textwidth]{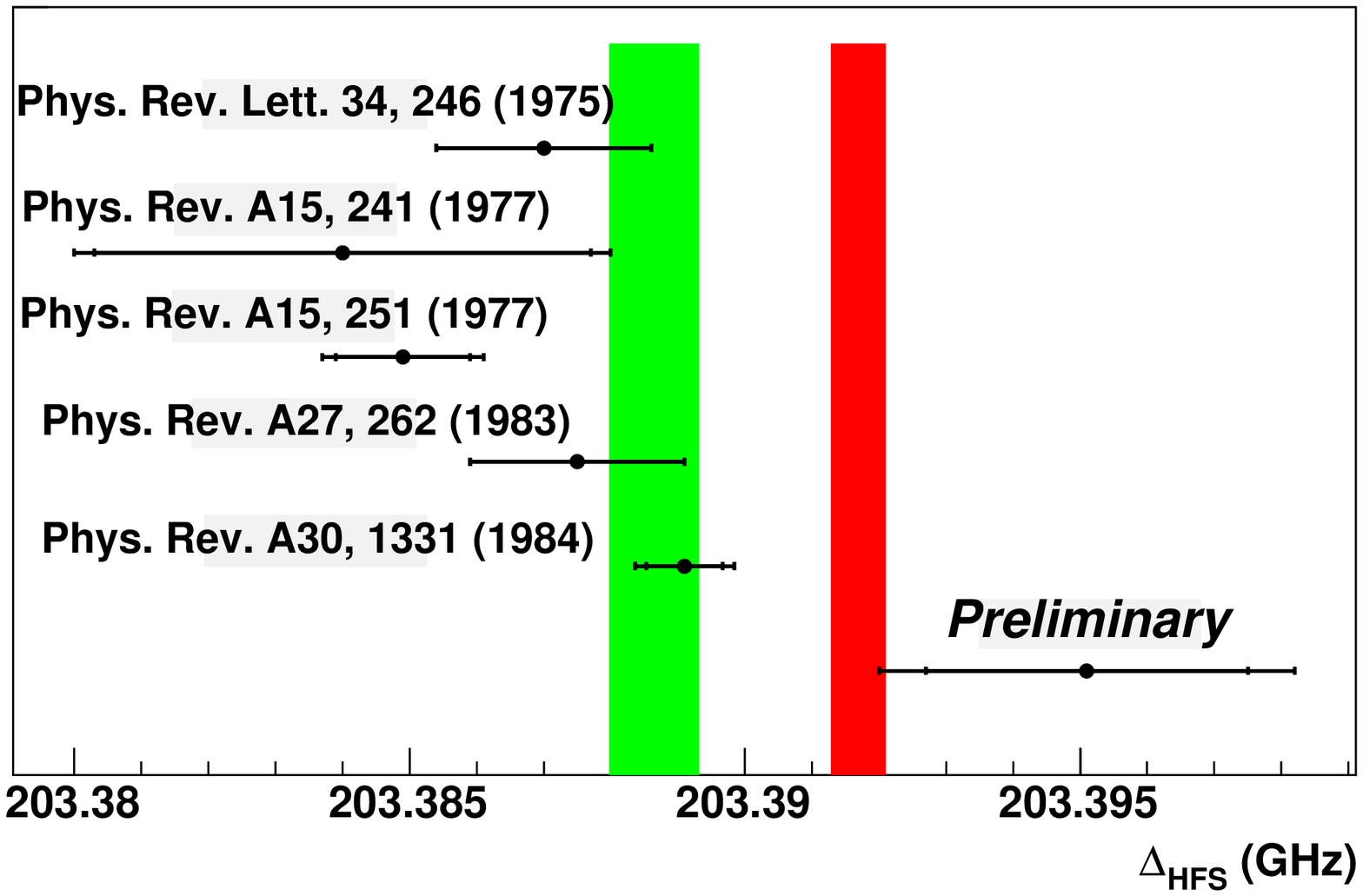}
\caption{Our current result in historical plot of $\Delta _{\mathrm{HFS}}$.}
\label{fig:current}
\end{figure}

The current result is
\begin{equation}
\Delta _{ \mathrm{HFS} } = 203.395\,1 \pm 0.002\,4 (\mathrm{stat.}, 12\,\mathrm{ppm}) \pm 
0.001\,9 (\mathrm{sys.}, 9.5\,\mathrm{ppm}) \, \mathrm{GHz} \, ,
\end{equation}
which is shown in Fig.~\ref{fig:current} compared with previous experiments and theoretical calculation.

\begin{table}
\caption{Summary of systematic errors of our current result.}
\label{tab:systematicerrors}
\begin{tabular}{lr}
\hline \noalign{\smallskip}
Source & Errors in $\Delta _{\mathrm{HFS}}$ (ppm) \\
\noalign{\smallskip}\hline\noalign{\smallskip}
{\it Magnetic Field:} & \\
~~Non-uniformity & 1.8 \\
~~Offset and reproducibility & 1.0 \\
~~NMR measurement & 1.0 \\
{\it Detection Efficiency:} & \\
~~Estimation using simulation & 7.0 \\
{\it Material Effect:} & \\
~~Ps thermalization & 3.0 \\
{\it RF System:} & \\
~~RF power & 2.9 \\
~~$Q_L$ value of RF cavity & 4.3 \\
~~RF frequency & 1.0 \\
\noalign{\smallskip}\hline\noalign{\smallskip}
Quadrature sum & 9.5 \\
\noalign{\smallskip}\hline
\end{tabular}
\end{table}

\section{Prospects}
\begin{enumerate}
\item {\it Material Effect} ~Currently we assumed that $\Delta_{\mathrm{HFS}}$ depends on 
gas density linearly. If the unthermalized Ps contribution is large, the dependence becomes non-linear. 
According to the previous measurement of the momentum-transfer cross section~\cite{NETSUKA-5}, 
unthermalized effect is 
estimated to be less than 3\,ppm with i-C$_4$H$_{10}$ gas. We are now precisely measuring the Ps 
thermalization using the same technique as 
ref.~\cite{KATAOKA,KATAOKA-D,JINNAI,JINNAI-D,ASAI,ASAI-D} (pick-off), whose technique is 
different from that of ref.~\cite{NETSUKA-5} (Doppler-broadening spectroscopy).
\item {\it RF System} ~We will carefully control the experimental environment, especially the temperature, 
which will reduce the uncertainty.
\item {\it Detection efficiency} ~Currently it is estimated by Monte Carlo simulation. 
It will be carefully studied and will be estimated by real data, which also will reduce the uncertainty.
\item {\it Statistics} ~12\,ppm has been obtained. By measuring more efficient points of gas density and 
magnetic field strengths, a measurement with a precision of $O$(ppm) is expected within a year.
\end{enumerate}

\section{Conclusion}
\label{sec:conclusion}
A new experiment to measure the Ps-HFS which reduces possible common uncertainties in previous experiments has been 
constructed.
The current result of $\Delta_{\mathrm{HFS}} = 203.395\,1 \pm 0.002\,4 (\mathrm{stat.}, 12\,\mathrm{ppm}) 
\pm 0.001\,9 (\mathrm{sys.}, 9.5\,\mathrm{ppm})\,\mathrm{GHz}$ has been obtained so far.
A new result with an accuracy of $O$(ppm) will be obtained within a year which will be an independent check 
of the discrepancy.


\bibliographystyle{spphys}       
\bibliography{LEAP2011-ishida}   

%
%

\end{document}